\documentclass[aip,jcp,reprint,noshowkeys,superscriptaddress]{revtex4-2}
\usepackage{graphicx,dcolumn,bm,xcolor,microtype,multirow,amscd,amsmath,amssymb,amsfonts,physics,wrapfig,bbold,siunitx,xspace}
\usepackage[version=4]{mhchem}

\usepackage[utf8]{inputenc}
\usepackage[T1]{fontenc}

\usepackage{hyperref}
\hypersetup{
    colorlinks,
    linkcolor={red!50!black},
    citecolor={red!70!black},
    urlcolor={red!80!black}
}

\usepackage{listings}
\definecolor{codegreen}{rgb}{0.58,0.4,0.2}
\definecolor{codegray}{rgb}{0.5,0.5,0.5}
\definecolor{codepurple}{rgb}{0.25,0.35,0.55}
\definecolor{codeblue}{rgb}{0.30,0.60,0.8}
\definecolor{backcolour}{rgb}{0.98,0.98,0.98}
\definecolor{mygray}{rgb}{0.5,0.5,0.5}

\definecolor{sqred}{rgb}{0.85,0.1,0.1}
\definecolor{sqgreen}{rgb}{0.25,0.65,0.15}
\definecolor{sqorange}{rgb}{0.90,0.50,0.15}
\definecolor{sqblue}{rgb}{0.10,0.3,0.60}

\lstdefinestyle{mystyle}{
    backgroundcolor=\color{backcolour},
    commentstyle=\color{codegreen},
    keywordstyle=\color{codeblue},
    numberstyle=\tiny\color{codegray},
    stringstyle=\color{codepurple},
    basicstyle=\ttfamily\footnotesize,
    breakatwhitespace=false,
    breaklines=true,
    captionpos=b,
    keepspaces=true,
    numbers=left,
    numbersep=5pt,
    numberstyle=\ttfamily\tiny\color{mygray},
    showspaces=false,
    showstringspaces=false,
    showtabs=false,
    tabsize=2
  }

  \newcolumntype{d}{D{.}{.}{-1}}

\newcommand{\fnm}{\footnotemark}
\newcommand{\fnt}{\footnotetext}
\newcommand{\mc}{\multicolumn}
\newcommand{\mcc}[1]{\mc{1}{c}{#1}}
\newcommand{\SupInf}{\textcolor{sqred}{Supporting Information}\xspace}

\newcommand{\EFCI}{E_\text{FCI}}
\newcommand{\EexFCI}{E_\text{exFCI}}
\newcommand{\Ndet}{N_\text{det}}
\newcommand{\Evar}{E_\text{var}}
\newcommand{\EPT}{E_\text{PT2}}
  
\newcommand{\LCPQ}{Laboratoire de Chimie et Physique Quantiques (UMR 5626), Universit\'e de Toulouse, CNRS, UPS, France}

\begin{document}	

\title{How Useful Can Selected Configuration Interaction Be?}
\title{Go Green: Selected Configuration Interaction as a More Sustainable Alternative for High Accuracy}
\author{Pierre-Fran\c{c}ois \surname{Loos}}
	\email{loos@irsamc.ups-tlse.fr}
	\affiliation{\LCPQ}
\author{Yann \surname{Damour}}
	\affiliation{\LCPQ}
\author{Abdallah \surname{Ammar}}
	\affiliation{\LCPQ}
\author{Michel \surname{Caffarel}}
	\affiliation{\LCPQ}
\author{F\'abris \surname{Kossoski}}
	\affiliation{\LCPQ}
\author{Anthony \surname{Scemama}}
	\affiliation{\LCPQ}
	
\begin{abstract}
Recently, a new distributed implementation of the full configuration interaction (FCI) method has been reported [\href{https://doi.org/10.1021/acs.jctc.3c01190}{Gao et al.~\textit{J. Chem Theory Comput.} \textbf{2024}, \textit{20}, 1185}]. Thanks to a hybrid parallelization scheme, the authors were able to compute the exact energy of propane (\ce{C3H8}) in the minimal basis STO-3G. This formidable task involves handling an active space of 26 electrons in 23 orbitals or a Hilbert space of \SI{1.3d12} determinants. This is, by far, the largest FCI calculation reported to date.
Here, we illustrate how, from a general point of view, selected configuration interaction (SCI) can achieve microhartree accuracy at a fraction of the computational and memory cost, via a sparse exploration of the FCI space. The present SCI calculations are performed with the \textit{Configuration Interaction using a Perturbative Selection made Iteratively} (CIPSI) algorithm, as implemented in a determinant-driven way in the \textsc{quantum package} software.
The present study reinforces the common wisdom that among the exponentially large number of determinants in the FCI space, only a tiny fraction of them significantly contribute to the energy. 
More importantly, it demonstrates the feasibility of achieving comparable accuracy using more reasonable and sustainable computational resources, hence reducing the ever-growing carbon footprint of computational chemistry.
\bigskip
\begin{center}
	\boxed{\includegraphics[width=0.5\linewidth]{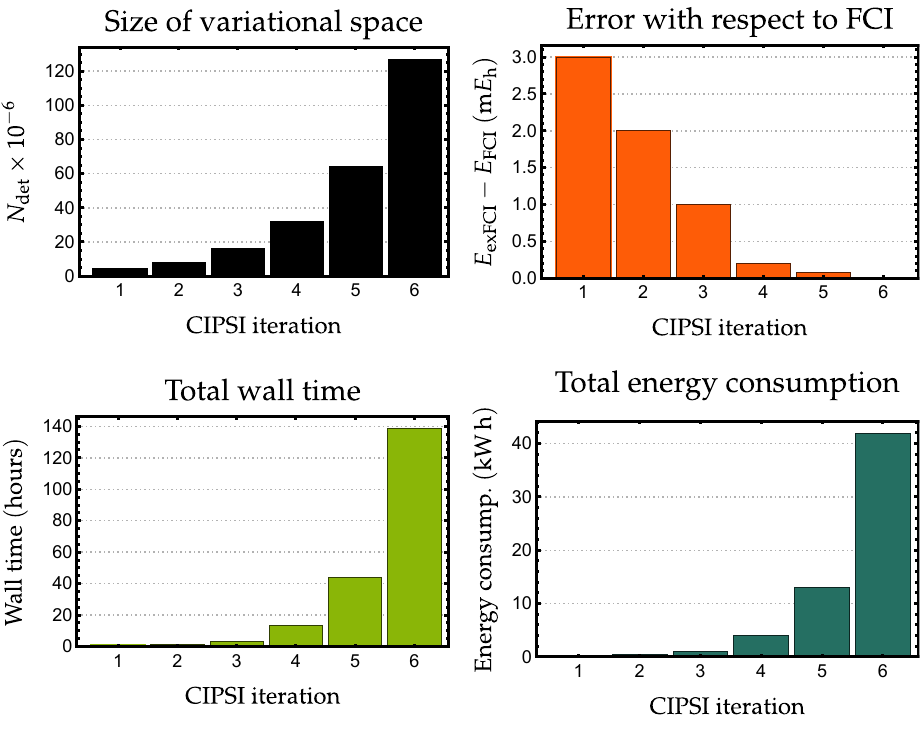}}
\end{center}
\bigskip
\end{abstract}

\maketitle

\section{Introduction}
\label{sec:introduction}

The full configuration interaction (FCI) method provides the exact solution of the Schr\"odinger equation in a given basis set, the number of electrons and orbitals defining the size of the Hilbert space. As such, it is often considered the ultimate answer when one wants to compare the performances of approximate methods. \cite{Loos2020a,Eriksen2020,Eriksen2021} However, the computational cost and memory requirement of a FCI calculation increases exponentially fast with the size of the system making it feasible on small systems only. FCI calculations on realistic molecular systems were achieved in the late 1980's, largely owing to the groundbreaking work of Knowles and Handy. \cite{Handy1980,Knowles1984,Knowles1989a,Knowles1989b} The significant milestone of surpassing the one-billion determinant barrier was accomplished in 1990 by Olsen et al. \cite{Olsen1990}  While there were several improvements during the first decade of the twenty-first century, \cite{Ansaloni2000,Gan2003,Rolik2008} the pace notably slowed thereafter. \cite{Fales2015,Vogiatzis2017} 

This slowdown can be ascribed partially to the emergence of alternative methodologies, notably selected configuration interaction (SCI) methods, which had lain dormant \cite{Bender1969,Whitten1969,Huron1973,Huron1973,Buenker1974} but underwent a resurgence in the early 2010's. \cite{Giner2013,Giner2015}
The essence of these methods lies in their sparse exploration of the Hilbert space, focusing only on the most energetically relevant determinants. This strategy stems from the realization that within the vast FCI space, only a tiny fraction of determinants significantly impact the energy.

Nowadays, SCI methods alongside related techniques like density-matrix renormalization group (DMRG) approaches \cite{White1992,White1993,Chan2011} and others, \cite{Eriksen2017,Eriksen2018,Eriksen2019a,Eriksen2019b,Motta2018,Lee2020,Xu2018,Xu2020,Magoulas2021,Gururangan2021} have become central to modern electronic structure theory. \cite{Loos2020a,Eriksen2020,Eriksen2021} Their main objective is to compute reference correlation and excitation energies in small molecular systems with exceptional accuracy, often rivaling FCI results. \cite{Eriksen2020,Loos2020b,Caffarel2016b,Holmes2017,Chien2018,Loos2018,Loos2019,Loos2020c,Veril2021}
Modern implementations of SCI encompass methodologies such as CIPSI (CI using a Perturbative Selection made Iteratively), \cite{Huron1973,Giner2013,Giner2015,Caffarel2016a,Garniron2017,Garniron2018,Garniron2019,Loos2020a,Loos2020b,Damour2021,Damour2023} adaptive sampling CI (ASCI), \cite{Schriber2016,Tubman2016,Tubman2018,Tubman2020} semistochastic heatbath CI (SHCI), \cite{Holmes2016,Holmes2017,Sharma2017,Smith2017,Chien2018,Li2018,Yao2020,Yao2021,Larsson2022} and iterative CI (iCI). \cite{Liu2014,Liu2016,Lei2017,Zhang2020,Zhang2021}
Stochastic CI techniques, like Monte Carlo CI (MCCI) \cite{Coe2018,Coe2022} and FCI quantum Monte Carlo (FCIQMC), \cite{Booth2009,Cleland2010,Blunt2015,Ghanem2019,Deustua2017,Deustua2018} adopt a similar approach, utilizing stochastic wave function representations to identify crucial determinants.

Recently, Gao et al. \cite{Gao2024} reported a new implementation of FCI using a hybrid parallelization scheme combining multiprocessing with MPI and multithreading with OpenMP. They were able to compute the exact energy of propane (\ce{C3H8}) in the STO-3G basis set. This calculation corresponds to a Hilbert space of size \SI{1.3d12} determinants. This formidable feat effectively breaks the one-trillion determinant barrier and is by far the largest FCI calculation reported to date.
According to Ref.~\onlinecite{Gao2024}, this calculation required running 512 processes on 256 nodes for a total wall time of 113.6 hours.
Their hardware consisted of nodes equipped with two Intel Xeon Gold 6148 CPUs, boasting a total of 
384 GB of memory and 40 logical cores per CPU, enabling the execution of 40 threads per process.
The most memory-consuming task requires an astonishing amount of 19 TB of memory.

In this study, our objective is to demonstrate that SCI calculations can be conducted routinely with significantly lower computational resources while still producing energies that closely match those obtained through FCI, thus contributing to the reduction of the ever-growing carbon footprint associated with computational chemistry.

\begin{table*}
  \caption{Convergence of the SCI ($\Evar$), SCI+PT2 ($\Evar + \EPT$) and exFCI  ($\EexFCI$) energies of propane (\ce{C3H8}) computed in the STO-3G basis with respect to the number of determinants included in the variational space ($\Ndet$). These calculations are performed with energetically-optimized orbitals. The error with respect to the FCI value of the SCI ($\Delta_\text{SCI}$), SCI+PT2 ($\Delta_\text{SCI+PT2}$), and exFCI ($\Delta_\text{exFCI}$) energies are also reported. The standard errors associated with the 3-point linear fitting procedure performed to obtain the exFCI energies are reported in parenthesis.}
  \label{tab:C3H8}
  \begin{ruledtabular}
\begin{tabular}{rccccll}
\mcc{$\Ndet$}	&	\mc{2}{c}{SCI energy}		&	\mc{2}{c}{SCI+PT2 energy}	&	\mc{2}{c}{exFCI energy}	\\
				\cline{2-3} \cline{4-5} \cline{6-7}
				&	$\Evar$ (\si{\hartree})		&	$\Delta_\text{SCI}$ (\si{\milli\hartree})		&	$\Evar+\EPT$  (\si{\hartree})			&	$\Delta_\text{SCI+PT2}$ (\si{\milli\hartree})	&	$\EexFCI$  (\si{\hartree})			&	$\Delta_\text{exFCI}$ (\si{\milli\hartree})		\\	
\hline
\num{1907}		&	\num{-117.08210421}	&	\num{18.018}	&	\num{-117.09776132}		&	\num{2.361}	&	&	\\
\num{3877}		&	\num{-117.08733585}	&	\num{12.787}	&	\num{-117.09848322}		&	\num{1.639}	&	&	\\
\num{7756}		&	\num{-117.09094151}	&	\num{9.181}		&	\num{-117.09900385}		&	\num{1.119}	&	&	\\
\num{15517}		&	\num{-117.09361712}	&	\num{6.506}		&	\num{-117.09936972}		&	\num{0.753}	&	&	\\
\num{31044}		&	\num{-117.09558736}	&	\num{4.535}		&	\num{-117.09959581}		&	\num{0.527}	&	&	\\
\num{62113}		&	\num{-117.09702968}	&	\num{3.093}		&	\num{-117.09978179}		&	\num{0.341}	&	&	\\
\num{124570}	&	\num{-117.09799309}	&	\num{2.130}		&	\num{-117.09989333}		&	\num{0.229}	&	&	\\
\num{249144}	&	\num{-117.09875792}	&	\num{1.365}		&	\num{-117.09998264}		&	\num{0.140}	&	\num{-117.100 143 5(7)}		&	\num{-0.0208(7)}	\\
\num{498290}	&	\num{-117.09926539}	&	\num{0.857}		&	\num{-117.10004465}		&	\num{0.078}	&	\num{-117.100 149(3)}		&	\num{-0.026(3)}	\\
\num{996650}	&	\num{-117.09959916}	&	\num{0.524}		&	\num{-117.10007604}		&	\num{0.047}	&	\num{-117.100 139(9)}		&	\num{-0.016(9)}	\\
\num{1993314}	&	\num{-117.09981192}	&	\num{0.311}		&	\num{-117.10009621}		&	\num{0.026}	&	\num{-117.100 125 8(1)}		&	\num{-0.0031(1)}	\\
\num{3986707}	&	\num{-117.09993942}	&	\num{0.183}		&	\num{-117.10010796}		&	\num{0.015}	&	\num{-117.100 125 5(3)}		&	\num{-0.0028(3)}	\\
\num{7973418}	&	\num{-117.10001798}	&	\num{0.105}		&	\num{-117.10011498}		&	\num{0.0077}	&	\num{-117.100 124 8(2)}		&	\num{-0.0021(2)}	\\
\num{15946880}	&	\num{-117.10006598}	&	\num{0.057}		&	\num{-117.10011864}		&	\num{0.0040}	&	\num{-117.100 123 7(5)}		&	\num{-0.0010(5)}	\\
\num{31893835}	&	\num{-117.10009352}	&	\num{0.029}		&	\num{-117.10012066}		&	\num{0.0020}	&	\num{-117.100 122 89(6)}	&	\num{-0.00021(6)}\\
\num{63788022}	&	\num{-117.10010827}	&	\num{0.014}		&	\num{-117.10012170}		&	\num{0.00098}	&	\num{-117.100 122 76(3)}	&	\num{-0.000 08(3)}\\
\num{126541040}	&	\num{-117.10011587}	&	\num{0.0068}	&	\num{-117.10012219}		&	\num{0.00049}	&	\num{-117.100 122 67(3)}	&	+\num{0.00001(3)}\\
\end{tabular}
  \end{ruledtabular}
\end{table*}

\begin{table}
  \caption{Energy of propane (\ce{C3H8}) computed at different levels of theory with the STO-3G basis. The error with respect to the FCI value is also reported.}
  \label{tab:energies}
  \begin{ruledtabular}
    \begin{tabular}{llc}
    	Method		&	Energy (\si{\hartree})	&	Error wrt FCI	\\	
	\hline
		FCI\fnm[1]		&	\num{-117.100 122 681 461}	&									\\	
		CCSD			&	\num{-117.098 767}			&	\SI{+1.355}{\milli\hartree}		\\	
		CCSD(T)			&	\num{-117.099 708}			&	\SI{+0.414}{\milli\hartree}		\\	
		CCSDT			&	\num{-117.099 942 158}		&	\SI{+0.181}{\milli\hartree}		\\	
		CCSDTQ			&	\num{-117.100 120 230}		&	\SI{+2.451}{\micro\hartree}		\\	
		SCI\fnm[2] 		&	\num{-117.100 093 52}		&	\SI{+0.029}{\milli\hartree}		\\	
		SCI+PT2\fnm[3] 	&	\num{-117.100 120 66}		&	\SI{+2.021}{\micro\hartree}		\\	
		exFCI\fnm[4] 	&	\num{-117.100 122 89(6)}	&	\SI{-0.21(6)}{\micro\hartree}	\\	
    \end{tabular}
  \end{ruledtabular}
  \fnt[1]{Reference \onlinecite{Gao2024}.}
  \fnt[2]{Variational energy obtained with $\Ndet = \num{32d6}$.}
  \fnt[3]{Perturbatively-corrected variational energy obtained with $\Ndet = \num{32d6}$.}
  \fnt[4]{Extrapolated FCI value obtained via a 3-point linear fit using $\Ndet = \num{32d6}$ as the largest variational space.}
\end{table}

\begin{table}
	\caption{Wall time, maximum memory consumption, total energy consumption, and error with respect to FCI of the exFCI energy for increasingly large CIPSI calculations performed on propane (\ce{C3H8}) in the STO-3G basis. The energy consumption represents the total consumption of the compute node measured by the BMC for the entire job duration (see main text).}
  \label{tab:conso}
  \begin{ruledtabular}
    \begin{tabular}{rcccc}
	    \mcc{$\Ndet$}	&	Wall time 	&	Memory		&	Energy 						&	Error 				\\
    					&	 (hh:mm)	&	consumption	&	consumption					&	wrt	FCI					\\
		\hline
        \num{2d6}		&	00:14	&	5.3 GB		&	\SI{74}{\watt\hour}			&	\SI{3}{\micro\hartree}		\\
        \num{4d6}		&	00:33	&	8.1 GB		&	\SI{176}{\watt\hour}		&	\SI{3}{\micro\hartree}		\\
        \num{8d6}		&	01:19	&	15 GB		&	\SI{438}{\watt\hour}		&	\SI{2}{\micro\hartree}		\\
        \num{16d6}		&	03:12	&	25 GB		&	\SI{1.1}{\kilo\watt\hour}	&	\SI{1}{\micro\hartree}		\\
        \num{32d6}		&	13:16	&	47 GB		&	\SI{4.1}{\kilo\watt\hour}	&	\SI{0.2}{\micro\hartree}	\\
        \num{64d6}		&	43:54	&	83 GB		&	\SI{13}{\kilo\watt\hour}	&	\SI{0.08}{\micro\hartree}	\\
        \num{127d6}		&	138:44	&	138 GB		&	\SI{42}{\kilo\watt\hour}	&	\SI{0.01}{\micro\hartree}	\\
    \end{tabular}
  \end{ruledtabular}
\end{table}

\section{Results and discussion}
\label{sec:res}

Our calculations are performed on a single node of the CALMIP supercomputer center (Toulouse, France), which has been in production since September 2018.
This node is a dual-socket Intel Skylake 6140 CPU@2.3 Ghz with 192 GB of memory for a total of 36 physical CPU cores.
The numbers given for the energy consumption of a calculation represent all the energy used by the compute node for the duration of the run.
These values were obtained from the database of the SLURM job scheduler, which was configured such that the \texttt{acct\_gather\_energy} plugin obtains the energy consumption from the Baseboard Management Controller (BMC) via the Intelligent Platform Management Interface (IPMI) protocol.

\begin{figure}
  \centering
  \includegraphics[width=0.8\linewidth]{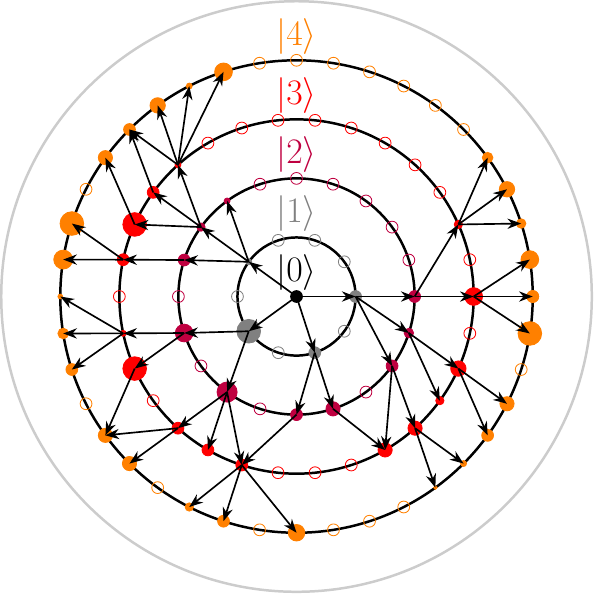}
  \caption{Schematic presentation of the growth of the variational space as a function of the CIPSI iterations. 
  Starting from a reference Slater determinant $\ket{0}$ built with a given set of orbitals, the variational space is enlarged systematically via the inclusion of the most energetically important determinants of higher excitation degrees (filled circles) to build larger multideterminant wave functions ($\ket{1}$, $\ket{2}$, $\ket{3}$, etc). In such a way, the Hilbert space is explored sparsely by leaving behind the determinants that do not contribute significantly to the energy (empty circles).
  }
   \label{fig:growth}
\end{figure}

The present SCI calculations have been carried out with the CIPSI algorithm. \cite{Huron1973,Giner2013,Giner2015,Garniron2017,Garniron2018,Garniron2019}
One of the main differences between conventional FCI implementations and the CIPSI implementation reported in Ref.~\onlinecite{Garniron2019} is that the former is integral-driven while the latter is determinant-driven.
(Note that the SHCI method, due to its different selection criteria, is also integral-driven. \cite{Holmes2016,Holmes2017,Sharma2017})
In a nutshell, CIPSI iteratively increases the size of the so-called variational space characterized by its (zeroth-order) variational energy $\Evar$ (see Fig.~\ref{fig:growth}).
As discussed further below, the variational energy (or SCI energy) converges slowly with respect to the number of determinants in the variational space $\Ndet$. To improve this convergence, $\Evar$ is corrected by its second-order perturbative energy, $\EPT$, computed within Epstein-Nesbet perturbation theory. \cite{Garniron2017}
The sum $\Evar+\EPT$ defines the SCI+PT2 energy.
Because there exists an approximate linear relationship between $\Evar$ and $\EPT$ when $\EPT$ is small enough, a linear extrapolation of $\Evar$ as $\EPT \to 0$ is performed to produce the final FCI estimate. \cite{Holmes2017,Burton2024} In the following, this extrapolated value is named exFCI.

As a first step, we perform a preliminary run where, in addition to iteratively increasing the size of 
the variational space, we further optimize the orbitals via the minimization of the variational energy. \cite{Levine2020,Yao2021,Damour2021} 
This orbital optimization procedure is performed at each CIPSI iteration up to \num{1.8d6} determinants.  
This run takes 2h56 of wall time, 4.9 GB of memory, and \SI{758}{\watt\hour} of energy consumption. 
The most memory-consuming task is the Davidson diagonalization of the largest variational space which requires 2.6 GB, while the most expensive PT2 calculation only requires 170 MB of memory.
(The orbital optimization step could easily be made globally less expensive without altering the overall accuracy by stopping the calculation earlier.)
Then, a new CIPSI run is performed with these optimized orbitals.
In Table \ref{tab:C3H8}, we report the evolution of the SCI ($\Evar$), SCI+PT2 ($\Evar+\EPT$), and exFCI ($\EexFCI$) energies as functions of $\Ndet$.
The errors with respect to the FCI energy, $\EFCI$, of Gao et al.~are also reported. \cite{Gao2024}
The exFCI values are obtained via a 3-point linear extrapolation of the variational energy using the three smallest $\EPT$ available at a given stage.

\begin{figure}
  \centering
  \includegraphics[width=0.8\linewidth]{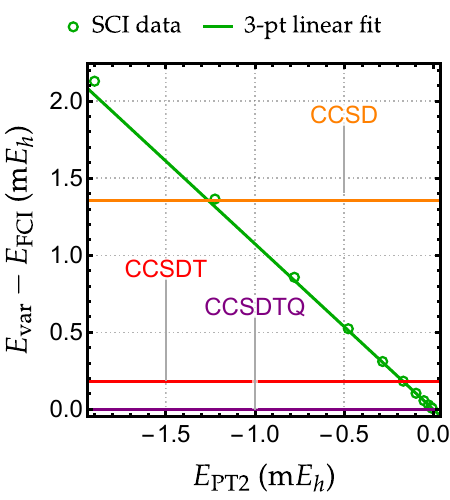}
  \caption{Variational energy as a function of the second-order perturbative correction for the SCI calculations performed on propane (\ce{C3H8}) in the STO-3G basis and with energetically-optimized orbitals.
  The 3-point linear fit is based on the three largest variational wave functions reported in Table \ref{tab:C3H8}.}
   \label{fig:extrap}
\end{figure}

The extrapolation procedure of $\Evar$ as a function of $\EPT$ is illustrated in Fig.~\ref{fig:extrap} for the largest variational space ($\Ndet = \num{127d6}$).
For the sake of comparison, the CCSD, CCSD(T), CCSDT, and CCSDTQ energies, computed with \textsc{cfour}, \cite{Matthews2020} are also reported in Fig.~\ref{fig:extrap} and Table \ref{tab:energies}. 
While the CCSD, CCSD(T), and CCSDT calculations require only a few seconds of wall time and a minimal amount of memory, the CCSDTQ calculation is more expensive as it requires 42 minutes of wall time and 508 MB of memory for an energy consumption of \SI{192}{\watt\hour}.
The CCSDTQ energy and the SCI+PT2 energy obtained with $\Ndet = \num{32d6}$ have similar accuracy and almost reach microhartree accuracy. Sub-microhartree accuracy can be obtained via extrapolation, the exFCI energy having an error of only \SI{-0.21(6)}{\micro\hartree} compared to FCI.

Table \ref{tab:conso} presents the wall time, maximum memory usage, and total energy consumption for increasingly larger CIPSI calculations. For large variational space, the Davidson diagonalization performed to compute the variational energy \cite{Garniron2019} is the most memory-consuming task while the semistochastic calculation of the second-order perturbative correction \cite{Garniron2017} is the most CPU-intensive task.
The accuracy of the exFCI energy (with respect to the FCI value) is also reported. Achieving microhartree accuracy necessitates approximately \num{16d6} determinants, 25 GB of RAM, and slightly over 3 hours of wall time, totaling approximately \SI{1.1}{\kilo\watt\hour}. Comparatively, a CIPSI calculation employing $\Ndet = \num{32d6}$ demands twice the memory, four times the energy consumption, and an additional 10 hours of wall time to achieve an overall accuracy of \SI{0.2}{\micro\hartree}.

\section{Conclusion}
\label{sec:conclusion}
We believe that these numbers illustrate nicely that SCI calculations are a more eco-friendly alternative compared to FCI calculations, especially for achieving high accuracy and generating reference values for benchmarking other computational methods. 
As the variational space expands (see Table \ref{tab:conso}), SCI calculations become more resource-intensive both in terms of computational resources and energy consumption. For example, to perform a calculation up to \num{126d6} determinants, almost 6 days of wall time are required for an energy consumption of \SI{42}{\kilo\watt\hour}. Hence, there remains ample scope for further optimization and improvement.

\acknowledgements{
The authors would like to thank Nicolas Renon (CALMIP) for his technical support.
This project has received funding from the European Research Council (ERC) under the European Union's Horizon 2020 research and innovation programme (Grant agreement No.~863481).
Additionally, it was supported by the European Centre of Excellence in Exascale Computing (TREX), and has received funding from the European Union's Horizon 2020 --- Research and Innovation program --- under grant agreement no.~952165.
This work was granted access to the HPC resources of CALMIP supercomputing center under the allocation 2024-18005.}

\section*{Supporting Information}
See the \SupInf for the geometry of propane and output files associated with the CC calculations, orbital optimization, and CIPSI calculations. 
Files gathering information about wall time, memory usage, and energy consumption are also provided.

\section*{Data availability statement}
The data that supports the findings of this study are available within the article and its \SupInf.

\section*{References}

\bibliography{sci}

\end{document}